\documentclass[MSNbibl,nameyear,dvips]{arxstspdf}
\usepackage{flushend}
\usepackage{stfloats}

\volume{26}
\issue{1}
\pubyear{2011}
\firstpage{17}
\lastpage{18}
\doi{10.1214/11-STS337B}
\referstodoi{10.1214/10-STS337}

\begin{document}
\begin{frontmatter}

\title{Discussion of ``Statistical Inference: The~Big Picture'' by R. E. Kass}
\runtitle{Discussion}
\pdftitle{Discussion of Statistical Inference: The Big Picture by R. E. Kass}

\begin{aug}
\author{\fnms{Hal} \snm{Stern}\corref{}\ead[label=e1]{sternh@uci.edu}}

\runauthor{H. Stern}

\affiliation{University of California, Irvine}

\address{Hal Stern is Ted and Janice Smith Family Foundation Dean of the Donald Bren School
of Information and Computer Sciences and Professor, Department of Statistics, University of California,
Irvine, 2216 Bren Hall, Irvine, California 92697, USA \printead{e1}.}

\end{aug}



\end{frontmatter}

Rob Kass presents a fascinating vision of a
``post''-Bayes/frequentist-controversy world in which practical utility
of statistical models is the guiding principle for statistical
inference.  I agree with Kass on many points.  In particular, Kass is
correct (in my opinion) when he notes that much modern statistical work
develops statistical models (the theoretical world) and asks whether
the models provide a~reasonable description or explanation of data (the
real world).  A recent example in my own collaborative work
(Scharenbroich et al., \citeyear{Schetal09}) builds a~storm tracking
model that combines subjective information from climate scientists
about storms in the eastern Pacific and historical data to develop a~%
probabilistic model that appears to fit data well.  A critical element
of this approach, as Kass notes, is that we understand the assumptions
that underlie our statistical model and, equally important, that we
subject these assumptions to careful scrutiny.  I continue to find
posterior predictive model checks (Gelman, Meng and Stern,
\citeyear{GelMenSte96}) especially helpful for asses\-sing model fit.

Of course, this would not be a particularly interesting discussion if
it focused on points of agreement.  I believe that Kass's proposed
``big picture'' fails at one key goal that we should have for such a
picture---it does not easily illustrate one of the key concepts of the
field, the art of generalizing from sample data to larger populations.
I argue below that the ``old'' big picture (Kass's Figure~3) still has
great value for me and for the field.  I also speculate a bit about
pragmatism as a foundation on which to build a training program for
statisticians.

\section*{In Defense of the ``Old'' Big Picture}

My main disagreement with Kass concerns his dissatisfaction with his
Figure~3 and the story that it tells.  According to Kass, the figure,
which describes inference as drawing conclusions about a population
from a sample of that population, ``is not a good general description
of statistical inference''; he prefers his Figure~1.  When it comes to
teaching introductory students, I~much prefer the old figure. The
statistical or quantitative literacy that I would love for my
introductory students to develop (and bring into the world with them)
does emphasize statistical inference as the process of learning about
populations from samples. Understanding the importance of the inference
question posed in this way will help non-statisticians ask whether a
study of memory in college sophomore psychology students provides
sufficient insight to allow one to generalize to the U.S. population as
a whole or whether a medical study associating a particular risk factor
with disease is based on a sufficiently representative sample.  When I~meet
with scientists on campus the starting point is not the
methodology but the scientific question and how to design a study that
will inform about that question.  The question of how to obtain
representative data is an important one and many studies suffer when
insufficient attention is paid to this basic point at the start of a
study. When I am asked about statistics by people outside the
University, ranging from middle school and high school students to my
in-laws and the occasional taxicab driver, I tell them about how we use
samples to learn about populations rather than about building
theoretical models of the real world.

The ``old'' big picture (Figure 3) is also an accurate reflection of
the world of survey sampling which plays a major role in the collection
of data that drives public policy. Survey sampling may not be a major
part of the statistical toolkit for the scientific collaborations
discussed by Kass but it remains a critical function of the discipline.
I would prefer future politicians learn about survey sampling and
statistical inference from the traditional picture than about
alternative binary regression models from the new big picture. Just
this summer the Canadian government proposed making their Census long
form optional---I would sure like for people to easily grasp why that
is problematic.  I believe they would see the problem from Figure 3 or
at a~minimum that the problem is easily described by referring to that
figure. Is the problem with optional survey response clear in Figure~1
(or even Figure 4)? I suppose the point could be made more general:
Kass's Figure~1 does not talk at all about where the data (the sample)
come from.  We know this is a~critical question.

Kass is correct in pointing out that the population/sample picture is
limiting.  There are many situations for which that big picture fails.
It is hard to tell a compelling regression story with that picture or
to address the Hecht et al. psychophysical experiment featured by Kass.
It is at this point that the statistical models that inhabit Kass's
theoretical world enter as a very useful way to proceed. When the time
comes in my introductory course that the old picture fails to be
relevant I introduce the concepts in Kass's Figure~1.  In fact, I like
Kass's Figure~1 and can easily imagine integrating it as the ``second''
big picture that I show my students. One can even point out to more
advanced students (statistics majors, statistics graduate students)
that in the majority of modern interdisciplinary scientific
collaborations it is the ``new'' big picture that reflects how we
proceed once we have collected the data.

\section*{Pragmatism as a Foundation}

Beyond my concerns about whether pragmatism is appropriate for
introductory students and for teaching basic quantitative literacy, I
also wonder what implications statistical pragmatism has for our
graduate training programs. Would we teach Bayesian inference as merely
a set of tools for the pragmatist to draw upon when appropriate?  How
much time should we spend talking about subjective prior distributions?
Although Kass starts his abstract by noting that ``Statistics has moved
beyond the fre\-quentist--Bayesian controversies of the past,'' I suspect
there might be considerable disagreement about curricular issues such
as these.  I~worry more broadly that pragmatism might appear to
reinforce the notion of statistics as a set of techniques that we
``pull off the shelf'' when confronted with a data set of a~particular
type.  I certainly do not believe that is the intent of the philosophy
described here; in fact I am quite certain that Kass is not in favor of
such an approach. My question then is how do we develop students into
the kind of science-based statistical pragmatists that Kass would like
to see.  I do not see pragmatism itself as providing us with the
prescription for how to get there.  Indeed, Kass's pragmatism seems to
be a fairly evolved state for a statistician; it seems to require a
clear understanding of the various competing foundational arguments
that have preceded it historically.

\section*{Conclusion}

Statistical pragmatism appears to me to be an accurate description of
the practices of many modern statisticians. In that regard I appreciate
Rob Kass's contribution to starting a discussion about what we mean by
statistical pragmatism and what its implications are for teaching
introductory students and graduate students about statistical
inference. I am concerned that pragmatism as presented here fails to
get at key points regarding data collection and sampling that are
essential to both professional statisticians' and the general
population's understanding of inference.

\section*{Acknowledgments}

This work was supported in part by NSF Grant AGS1025374.



\begin{thebibliography}{3}

\bibitem[\protect\citeauthoryear{Gelman, Meng and Stern}{1996}]{GelMenSte96}
\begin{barticle}[mr]
\bauthor{\bsnm{Gelman},~\bfnm{Andrew}\binits{A.}},
  \bauthor{\bsnm{Meng},~\bfnm{Xiao-Li}\binits{X.-L.}} \AND
  \bauthor{\bsnm{Stern},~\bfnm{Hal}\binits{H.}}
(\byear{1996}).
\btitle{Posterior predictive assessment of model fitness via realized
  discrepancies (with discussion)}.
\bjournal{Statist. Sinica}
\bvolume{6}
\bpages{733--807}.
\bid{issn={1017-0405}, mr={1422404}}
\bptnote{check related}%
\end{barticle}
\endbibitem

\bibitem[\protect\citeauthoryear{Scharenbroich et~al.}{2009}]{Schetal09}
\begin{barticle}[auto:STB|2011-03-03|12:04:44]
\bauthor{\bsnm{Scharenbroich},~\bfnm{L.}\binits{L.}},
  \bauthor{\bsnm{Magnusdottir},~\bfnm{G.}\binits{G.}},
  \bauthor{\bsnm{Smyth},~\bfnm{P.}\binits{P.}},
  \bauthor{\bsnm{Stern},~\bfnm{H.}\binits{H.}} \AND
  \bauthor{\bsnm{Wang},~\bfnm{C.~C.}\binits{C.~C.}}
(\byear{2009}).
\btitle{A Bayesian framework for storm tracking using a hidden-state
  representation}.
\bjournal{Mon. Weather Rev.}
\bvolume{138}
\bpages{2132--2148}.
\end{barticle}
\endbibitem

\end{thebibliography}
\end{document}